 \documentclass[final,3p,times]{elsarticle}

\usepackage{graphicx}

\usepackage{amssymb}

\biboptions{compress}

\begin{document}

\begin{frontmatter}

\title{Robustness of quantum discord to sudden death in NMR}

 \author{Jianwei Xu\corref{cor1}\fnref{1}}
  \address[1]{Key Laboratory for Radiation Physics and Technology, Institute of Nuclear Science and Technology, Sichuan University, Chengdu 610065, China}
 \cortext[cor1]{Corresponding author. Tel.: +86 28 85412379; fax: +86 28 8541 0252}
 \ead{xxujianwei@yahoo.cn}

\author[2]{Qihui Chen}
\address[2]{Physics department, Sichuan University, Chengdu 610065, China}

\begin{abstract}
We investigate the dynamics of entanglement and quantum discord of
two qubits in liquid state homonuclear NMR. Applying a phenomenological description for NMR under relaxation process, and taking a group of typical parameters of NMR, we show that when a zero initial state $|00\rangle $
experiences a relaxation process, its entanglement disappears completely after a
sequence of so-called sudden deaths and revivals, while the quantum discord retains remarkable values after a sequence of oscillations. That is to say, the quantum discord is more robust than entanglement.

\end{abstract}


\end{frontmatter}

\section{Introduction}
Entanglement is one of the most striking features for our deeper
understanding of the quantum world, and is widely seen as the main resource
for quantum information processing \cite{Nielson2000,Horodecki2009}. However,
entanglement is not the only type of useful correlation. It has been found that entanglement is not necessary for deterministic quantum computation with one pure qubit (DQC1) \cite{Knill1998,Datta2008,Lanyon2008}, but another quantum
correlation, introduced by \cite{Ollivier2001,Henderson2001}, called quantum
discord (QD), is responsible for the computational efficiency of DQC1.
Besides, QD has also been used for some studies such as quantum phase transition
\cite{Dillenschneider2008} and the process of Grover search\cite{Cui2009}. Recently,
the dynamics of QD has attracted more attentions, such as the robustness of QD to sudden
death \cite{Werlang2009}, QD in Heisenberg models \cite{Werlang2010} and Non-Markovian
effect of QD \cite{Wang2010}.

Nuclear magnetic resonance (NMR) has been used to demonstrate basic concepts
of quantum information processing, and has been leading the field both in
terms of number of qubits and control techniques
\cite{Vandersypen2005,Oliveira2007}. The studies of dynamics characteristics and
the implementations for quantum information processing of NMR systems have also
received a great deal of attentions in recent years
\cite{Fahmy2008,Furman2008,Rufeil-Fiori2009,Ota2009,Zhang2009,Soares-Pinto2010}.

Due to the fundamental and practical significance, we wish to investigate
the dynamics behaviors of entanglement and QD of two qubits in NMR systems.
The rest of this paper is organized as follows. In Sec.2, we  present a brief review
of the entanglement, QD and liquid state two qubits NMR systems. In Sec.3,
we investigate the dynamics of entanglement and quantum discord of two
qubits NMR without relaxation. In Sec.4, applying a phenomenological method, we investigate the dynamics of
entanglement and quantum discord of two qubits NMR with relaxation. Sec.5 is
a brief summary.

\section{Entanglement,quantum discord, NMR systems}
We first recall some basics about entanglement, QD and NMR.

\subsection{Concurrence of two qubits states}
 Two quantum systems A and
B, described by the Hilbert spaces $H^A$ and $H^B$, and the composite system
AB is described by the Hilbert space $H^A$ $\otimes $ $H^B$.
 A state $\rho $ on $H^A$ $\otimes $ $H^B$ is
called a separable state if and only if it can be written as the form
\begin{eqnarray}
\rho =\sum_{i=1}^np_i\rho _i^A\otimes \rho _i^B,
\end{eqnarray}
where n is a positive integer, $p_i\geq 0$, $\sum_{i=1}^np_i=1$, $\{\rho
_i^A\}_{i=1}^n$ are density operators on $H^A$, $\{\rho _i^B\}_{i=1}^n$ are
density operators on $H^B$. A state is called an entangled state if it is
not a separable state. For two qubits, one of the most widely used
entanglement measures is the so-called concurrence, we will use it as the
quantifier in this paper. For two qubits state $\rho $, the concurrence of $\rho $ is
defined as \cite{Wootters1998}
\begin{eqnarray}
C(\rho )=\max \{0,\lambda _1-\lambda _2-\lambda _3-\lambda _4\},
\end{eqnarray}
where $\lambda _1$, $\lambda _2$, $\lambda _3$, $\lambda _4$ are the square
roots of the eigenvalues of the matrix $\rho \widetilde{\rho }$ in
decreasing order, and
\begin{eqnarray}
\widetilde{\rho }=(\sigma _y\otimes \sigma _y)\rho ^{*}(\sigma _y\otimes
\sigma _y),
\end{eqnarray}
$\sigma _y$ is the Pauli matrix, $\rho ^{*}$ is the complex conjugate of $\rho $. For convenience, in Eq.(2) we denote
\begin{eqnarray}
\overline{C}(\rho )=\lambda _1-\lambda _2-\lambda _3-\lambda _4,
\end{eqnarray}
and call it pseudo-concurrence.

\subsection{Quantum discord of two qubits states}
The quantum discord of $\rho $ on $H^A$ $\otimes $ $H^B$ (with respect to system A) is defined as
\cite{Ollivier2001,Henderson2001}
\begin{eqnarray}
D_A(\rho )=S(\rho ^A)-S(\rho ^{AB})+\inf_{\{|\alpha \rangle \langle \alpha |\otimes I\}_{\alpha =1}^{n_A}}\sum_{\alpha =1}^{n_A}p_\alpha S(\rho _\alpha ^B).
\end{eqnarray}
 In Eq.(5), $S(\cdot )$ is the entropy function, that is, e.g., $S(\rho ^A)=-tr_A(\rho ^A\log _2\rho ^A)$. $n_A=dimH^A$, inf takes all projective measurements $\{|\alpha \rangle \langle \alpha |\otimes I\}_{\alpha =1}^{n_A}$ on system A, that is, $\{|\alpha \rangle \}_{\alpha =1}^{n_A}$ is an arbitrary orthonormal basis for $H^A$. We use $I$ to denote the identity operators on $H^A$ or on $H^B$. $p_\alpha =tr_A(|\alpha \rangle \langle \alpha |\otimes I\rho |\alpha \rangle \langle \alpha |\otimes I)$, $\rho _\alpha ^B=|\alpha \rangle \langle \alpha |\otimes I\rho |\alpha \rangle \langle \alpha |\otimes I/p_\alpha $.

Quantum discord captures more correlation than entanglement, this
can been seen by Eq.(1) and the fact that \cite{Ollivier2001}
\begin{eqnarray}
D_A(\rho )=0\Longleftrightarrow \rho =\sum_{\alpha =1}^{n_A}p_\alpha |\alpha \rangle \langle \alpha |\otimes \rho _\alpha ^B,
\end{eqnarray}
where, $\{|\alpha \rangle \}_{\alpha =1}^{n_A}$ is an arbitrary orthonormal basis for $H^A$, $\{\rho _\alpha ^B\}_{\alpha =1}^{n_A}$ are density operators on $H^B$, $p_\alpha \geq 0$, $\sum_{\alpha =1}^{n_A}p_\alpha =1$. A state $\rho $ satisfying $D_A(\rho )=0$ is called zero quantum discord state or classical state. Comparing Eq.(1) with Eq.(6), it is obvious that a classical state must be a separable state.

Eq.(5) is difficult to optimize, even for two qubits states, up to now only few special states were
found to allow analytical solutions \cite{Luo2008,Ali2010}. A technical definition of
quantum discord, called geometrical measure, was introduced as \cite{Dakic2010},
\begin{eqnarray}
D_A^G(\rho )=\inf_\sigma tr[(\rho -\sigma )^2],
\end{eqnarray}
where inf takes all $\sigma $ that $D_A(\sigma )=0$. Evidently,
\begin{eqnarray}
D_A(\rho )=0\Longleftrightarrow D_A^G(\rho )=0.
\end{eqnarray}
One of the most elegant results about the definition of $D_A^G(\rho )$ is
that it allows analytical expressions for all two qubits states. More
specifically, for any two qubits state $\rho $ which can be written as
\begin{eqnarray}
\rho =\frac 14(I\otimes I+\sum_{i=1}^3x_i\sigma _i\otimes
I+\sum_{j=1}^3y_jI\otimes \sigma _j+\sum_{i,j=1}^3T_{ij}\sigma _i\otimes
\sigma _j),
\end{eqnarray}
then \cite{Dakic2010},
\begin{eqnarray}
D_A^G(\rho )=\frac 14(\sum_{i=1}^3x_i^2+\sum_{i,j=1}^3T_{ij}^2-\lambda
_{\max }).
\end{eqnarray}
Where, $\sigma _1=\sigma _x$, $\sigma _2=\sigma _y$, $\sigma _3=\sigma _z$,
are Pauli matrices, $\{x_i\}_{i=1}^3$, $\{y_j\}_{j=1}^3$, $%
\{T_{ij}\}_{i,j=1}^3$ are all real number sets. $\lambda _{\max }$ is the
largest eigenvalue of the matrix $xx^t+TT^t$, $x=(x_1,x_2,x_3)^t$, $T$ is
the matrix $(T_{ij})$, t means transpose. Also
\begin{eqnarray}
x_i=tr[\rho \sigma _i\otimes I],y_j=tr[\rho I\otimes \sigma
_j],T_{ij}=tr[\rho \sigma _i\otimes \sigma _j].
\end{eqnarray}
(The explicit expressions for $\{x_i\}_{i=1}^3$ and $%
\{T_{ij}\}_{i,j=1}^3$ by the elements of $\rho $, see Appendix A.)

\subsection{Two qubits in liquid state NMR}
The Hamiltonian of two qubits in NMR (we focus on the liquid state NMR) is well
described by \cite{Vandersypen2005}
\begin{eqnarray}
&& \ \ H(t)=H_{sys}+H_{rf}(t), \\
&& \ \ H_{sys}=-\frac{\hbar \omega _1}2\sigma _z\otimes I-\frac{\hbar \omega _2}%
2I\otimes \sigma _z+\hbar J\sigma _z\otimes \sigma _z,  \\
&&H_{rf}(t)=-\frac{\hbar g_1}2[\cos (\omega t)\sigma _x\otimes I-\sin (\omega
t)\sigma _y\otimes I]-\frac{\hbar g_2}2[\cos (\omega t)I\otimes \sigma _x-\sin (\omega
t)I\otimes \sigma _y].
\end{eqnarray}

In $H_{sys}$, $\omega _1=B_0\gamma _1$, $\omega _2=B_0\gamma _2$, are called
Lamor frequencies of two qubits (two nuclei), containing the chemical shifts. $%
B_0$ is the static magnetic field along z direction, $\gamma _1$, $\gamma _2$
are gyromagnetic ratios of two qubits. Typical values of $\omega _1$, $\omega _1$ are a few $10^{8}$Hz, and chemical shifts of a few $10^{3}$Hz to a few $10^{4}$Hz. J describes the spin-spin couplings
including direct dipole-dipole coupling and indirect through-bond coupling.
Spin-spin couplings are very small in NMR compared with the Lamor
frequencies, for instance, a few $10^{2}$Hz.

In $H_{rf}(t)$, $g_1=B_1\gamma _1$, $g_2=B_1\gamma _2$, $B_1$ is the applied
magnetic field rotating in x-y plane at frequency $\omega $, at or near
the Lamor frequencies $\omega _1$, $\omega _2$. Typical values of $g_1$, $g_2$ are up to $10^{5}$Hz.

\section{Dynamics of entanglement and quantum discord of two qubits in NMR without
relaxation}

We study the dynamics of two qubits NMR without
relaxation. When the evolution time is much shorter than the relaxation
time scales, the system can be approximated as an isolated system, and it evolves obeying the
Schr$\ddot{o}$dinger equation
\begin{eqnarray}
i\hbar \frac \partial {\partial t}|\Psi (t)\rangle =H(t)|\Psi (t)\rangle.
\end{eqnarray}
Notice that $H_{rf}(t)$ in Eq.(14) is time dependent, in order to cancel the time t in
Hamiltonian $H(t)$, we put
\begin{eqnarray}
|\Psi (t)\rangle =\exp [\frac{i\omega t}2(\sigma _z\otimes I+I\otimes
\sigma _z)]|\varphi (t)\rangle.
\end{eqnarray}
This transformation is widely used in NMR theory, sometimes called the method of rotating frame. With this transformation, we get

\begin{eqnarray}
&& \ \  i\hbar \frac \partial {\partial t}|\varphi (t)\rangle = \widetilde{H}%
|\varphi (t)\rangle, \\
&&\widetilde{H}= -\frac{\hbar (\omega _1-\omega )}2\sigma _z\otimes I-\frac{%
\hbar (\omega _2-\omega )}2I\otimes \sigma _z+\hbar J\sigma _z\otimes \sigma
_z -\frac{\hbar g_1}2\sigma _x\otimes I-\frac{\hbar g_2}2I\otimes \sigma _x, \\
&& \ \  |\Psi (0)\rangle = |\varphi (0)\rangle.
\end{eqnarray}
In above reductions, we have used the facts that for any real number $\lambda$,
\begin{eqnarray}
&& \ \ \ \ \ \ \ \ \  \ \ \ exp(i\lambda \sigma _i)=\cos (\lambda )I+i\sin (\lambda )\sigma _i, \\
&&exp(-i\lambda \sigma _z)\sigma _xexp(i\lambda \sigma _z)=\cos (2\lambda
)\sigma _x+\sin (2\lambda )\sigma _y, \\
&&exp(-i\lambda \sigma _z)\sigma _yexp(i\lambda \sigma _z)=\cos (2\lambda
)\sigma _y-\sin (2\lambda )\sigma _x.
\end{eqnarray}
Since $\widetilde{H}$ is time independent, together with Eq.(16) and Eq.(19), we get
\begin{eqnarray}
&&|\varphi (t)\rangle =\exp (\frac t{i\hbar }\widetilde{H})|\varphi
(0)\rangle,  \\
&&|\psi (t)\rangle =U(t)|\psi (0)\rangle,  \\
&&U(t)=\exp [\frac{i\omega t}2(\sigma _z\otimes I+I\otimes \sigma _z)]\exp
(\frac t{i\hbar }\widetilde{H}).
\end{eqnarray}
In matrix notation,
\begin{eqnarray}
\exp [\frac{i\omega t}2(\sigma _z\otimes I+I\otimes \sigma _z)]=\left(
\begin{array}{llll}
e^{i\omega t} & 0 & 0 & 0 \\
0 & 1 & 0 & 0 \\
0 & 0 & 1 & 0 \\
0 & 0 & 0 & e^{-i\omega t}
\end{array}
\right),
\end{eqnarray}
\begin{eqnarray}
\widetilde{H}=-\hbar \left(
\begin{array}{llll}
\frac{\omega _1+\omega _2}2-\omega -J & \frac{g_2}2 & \frac{g_1}2 & 0 \\
\frac{g_2}2 & \frac{\omega _1-\omega _2}2+J & 0 & \frac{g_1}2 \\
\frac{g_1}2 & 0 & -\frac{\omega _1-\omega _2}2+J & \frac{g_2}2 \\
0 & \frac{g_1}2 & \frac{g_2}2 & -\frac{\omega _1+\omega _2}2+\omega -J
\end{array}
\right).
\end{eqnarray}
Suppose the initial state is $|00\rangle $, that is

\begin{eqnarray}
\rho =\left(
\begin{array}{llll}
1 & 0 & 0 & 0 \\
0 & 0 & 0 & 0 \\
0 & 0 & 0 & 0 \\
0 & 0 & 0 & 0
\end{array}
\right)
\end{eqnarray}
then
\begin{eqnarray}
C(\rho )=0,  \ \ \ \ \ \
D_A^G(\rho )=0.
\end{eqnarray}
After time t, $\rho $ will evolve to $\rho (t)$ as
\begin{eqnarray}
\rho (t)=U(t)\rho U^{\dagger }(t),
\end{eqnarray}
where $\dagger $ means Hermitian conjugate. Then $C(\rho (t))$, $D_A^G(\rho
(t))$ can be calculated by Eq.(2) and Eq.(10).

Fig.1 shows $\overline{C}(\rho (t))$ and $D_A^G(\rho (t))$ at  $t=10^{-5}s$,
$t=10^{-3}s$, $t=10^{-1}s$. Where we take the parameters of NMR as
$\frac{\omega _1+\omega _2}2=3\times 10^8Hz$,
$\frac{\omega _1-\omega _2}2=10^4Hz$,
$J=3\times 10^2Hz$,
$\frac{g_1}2=5\times 10^4Hz$,
$\frac{g_2}2=5\times 10^4Hz$.
From Fig.1 we see that, only near the frequency $\frac{\omega _1+\omega _2}2=3\times 10^8Hz$,
 entanglement and quantum discord are
remarkably generated and evolved.
 It is a common phenomenon in NMR that
the applied magnetic $B_{1}$ can effectively influence or control the nuclear spins at or near the resonance frequencies.

Fig.2 shows  $\overline{C}(\rho (t))$ and $D_A^G(\rho (t))$ at the
frequency $\frac{\omega _1+\omega _2}2=3\times 10^8Hz$, and we take the parameters of NMR as in Fig.1.
From Fig.1 and Fig.2 we see that the dynamical behaviors of
entanglement and quantum discord are similar in the evolution where we ignore the relaxation.

 \begin{figure}[h]
 \includegraphics[width=5cm]{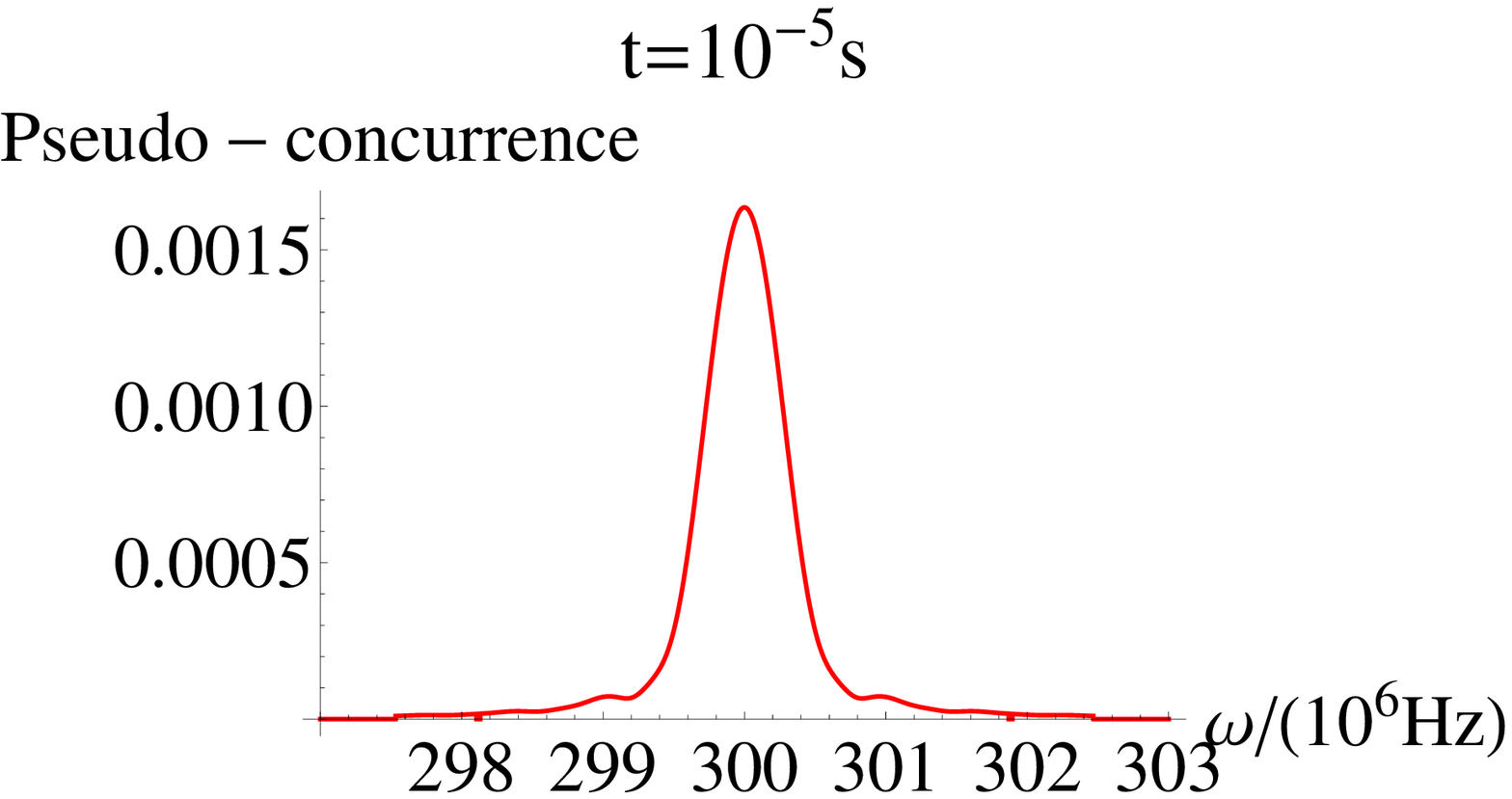}
 \includegraphics[width=5cm]{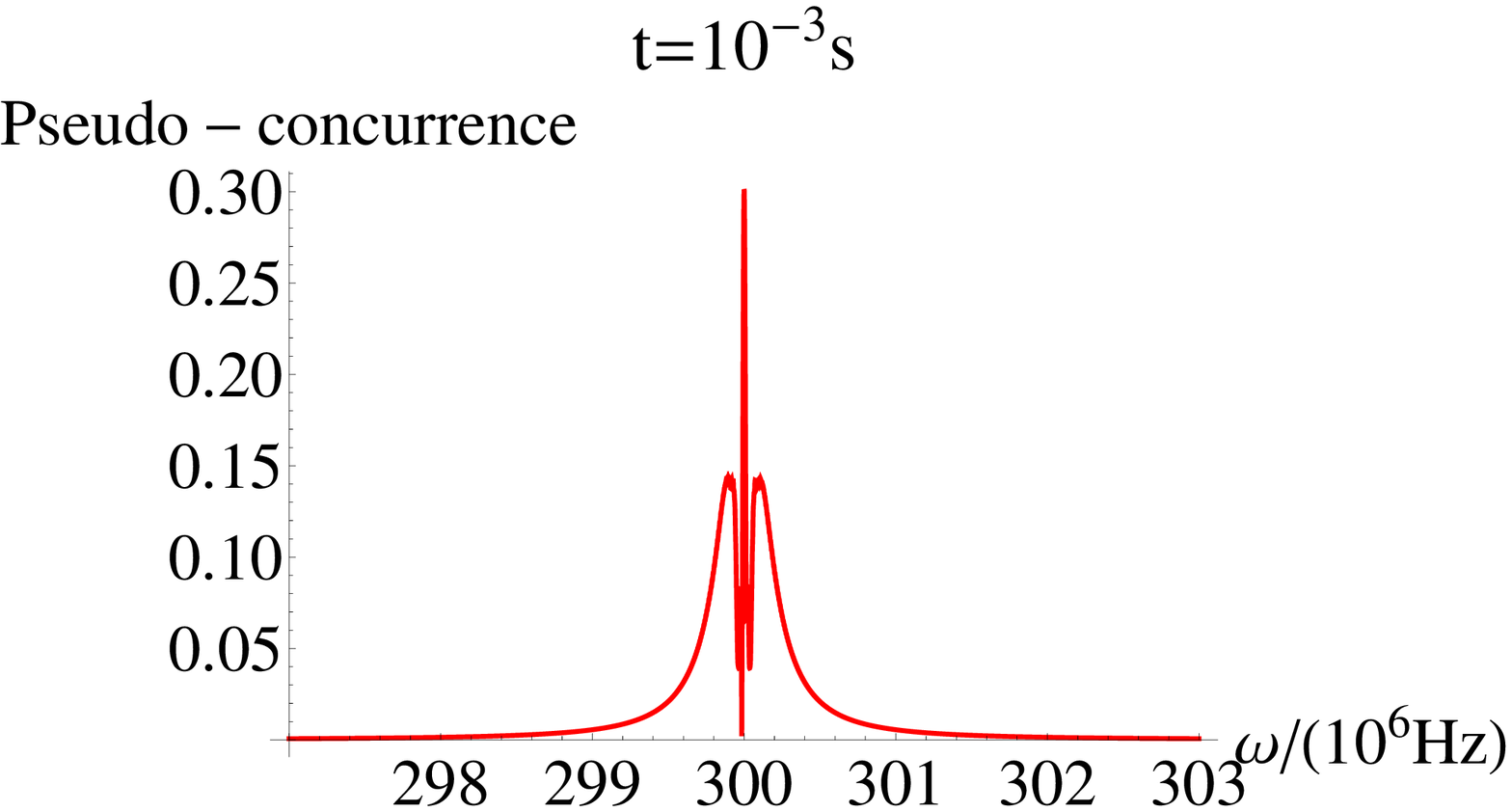}
 \includegraphics[width=5cm]{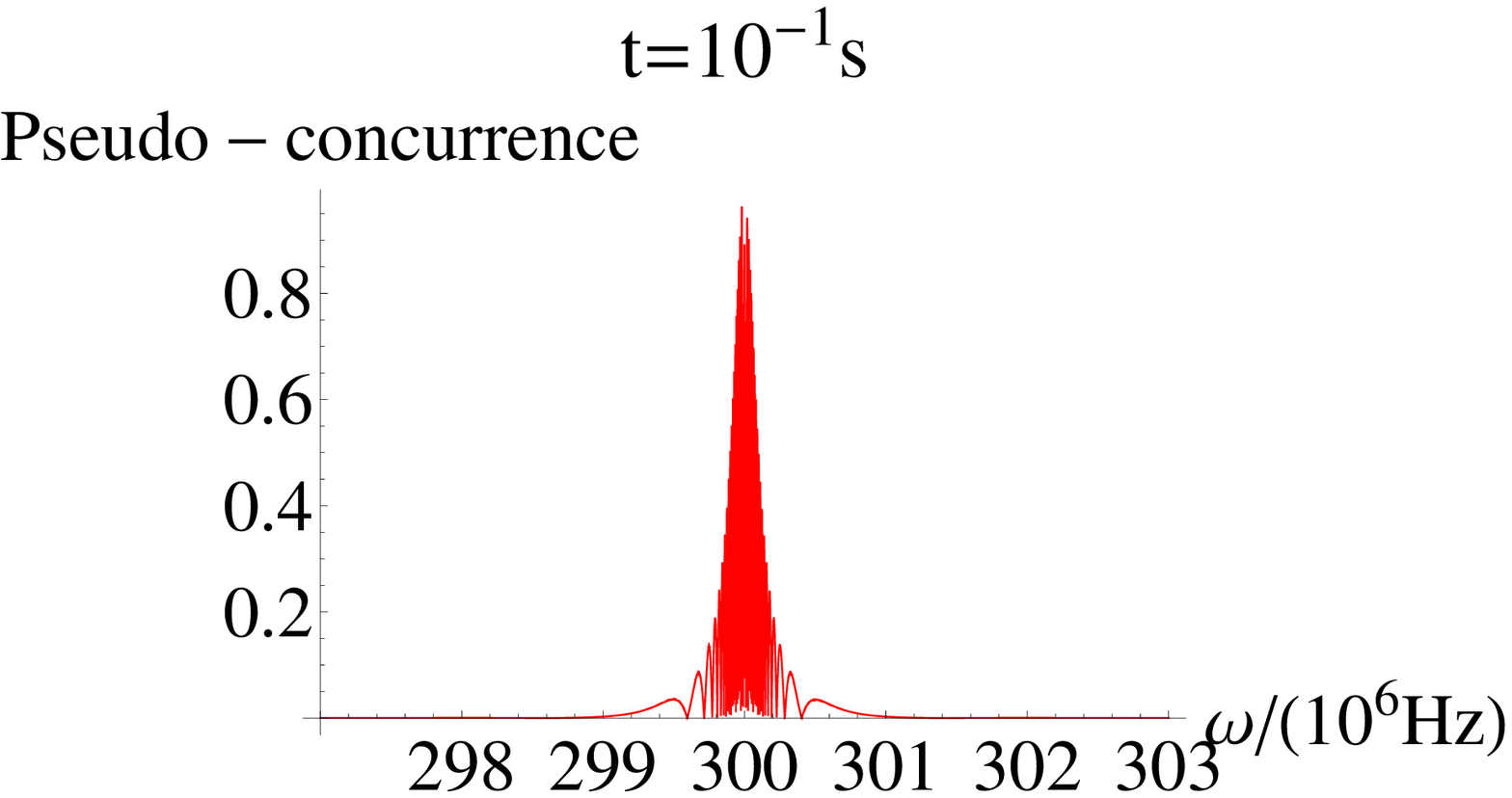}
 \end{figure}

 \begin{figure}[h]
 \includegraphics[width=5cm,trim=-2 0 0 0]{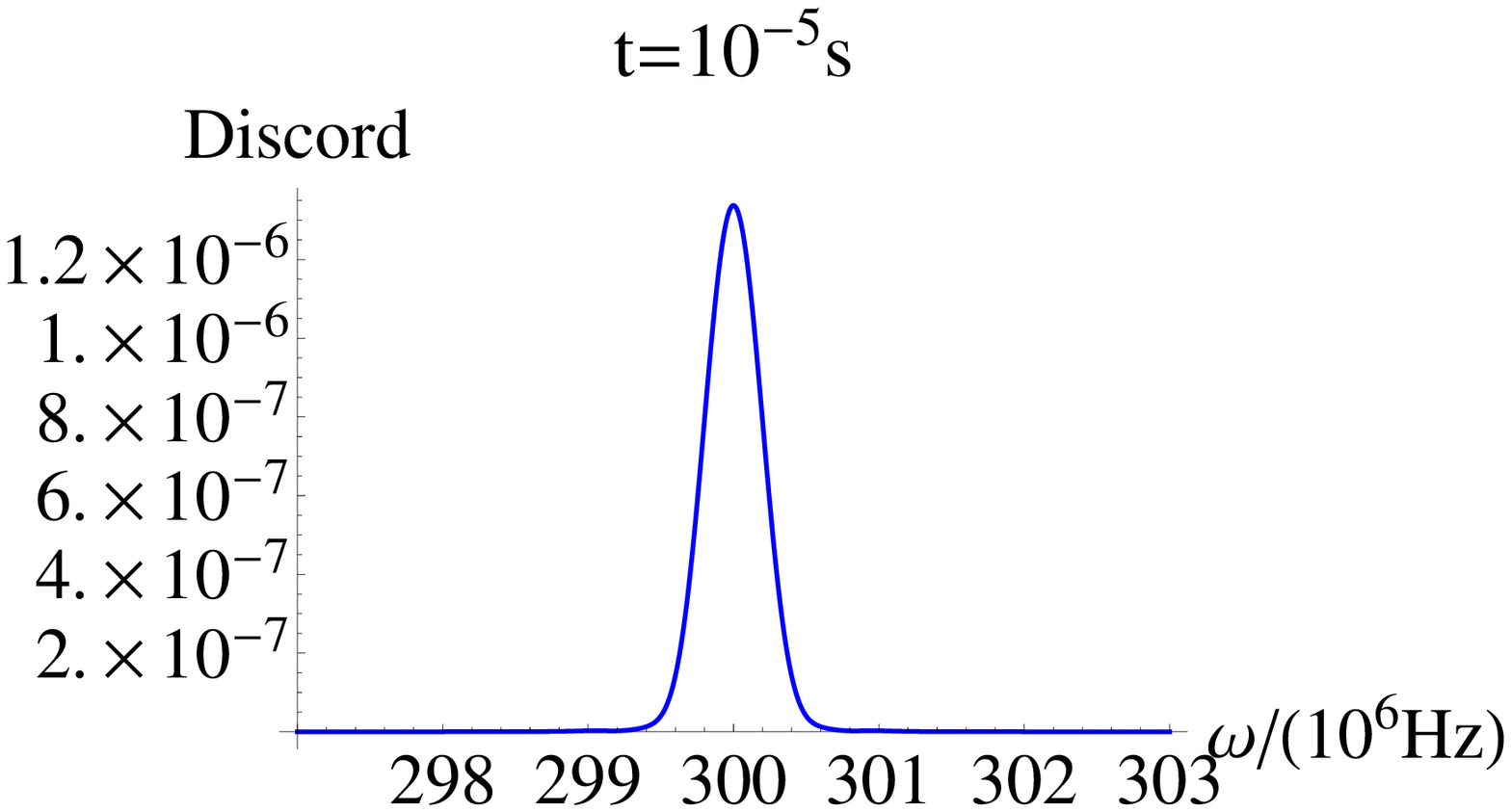} \ \ \ \
 \includegraphics[width=5cm]{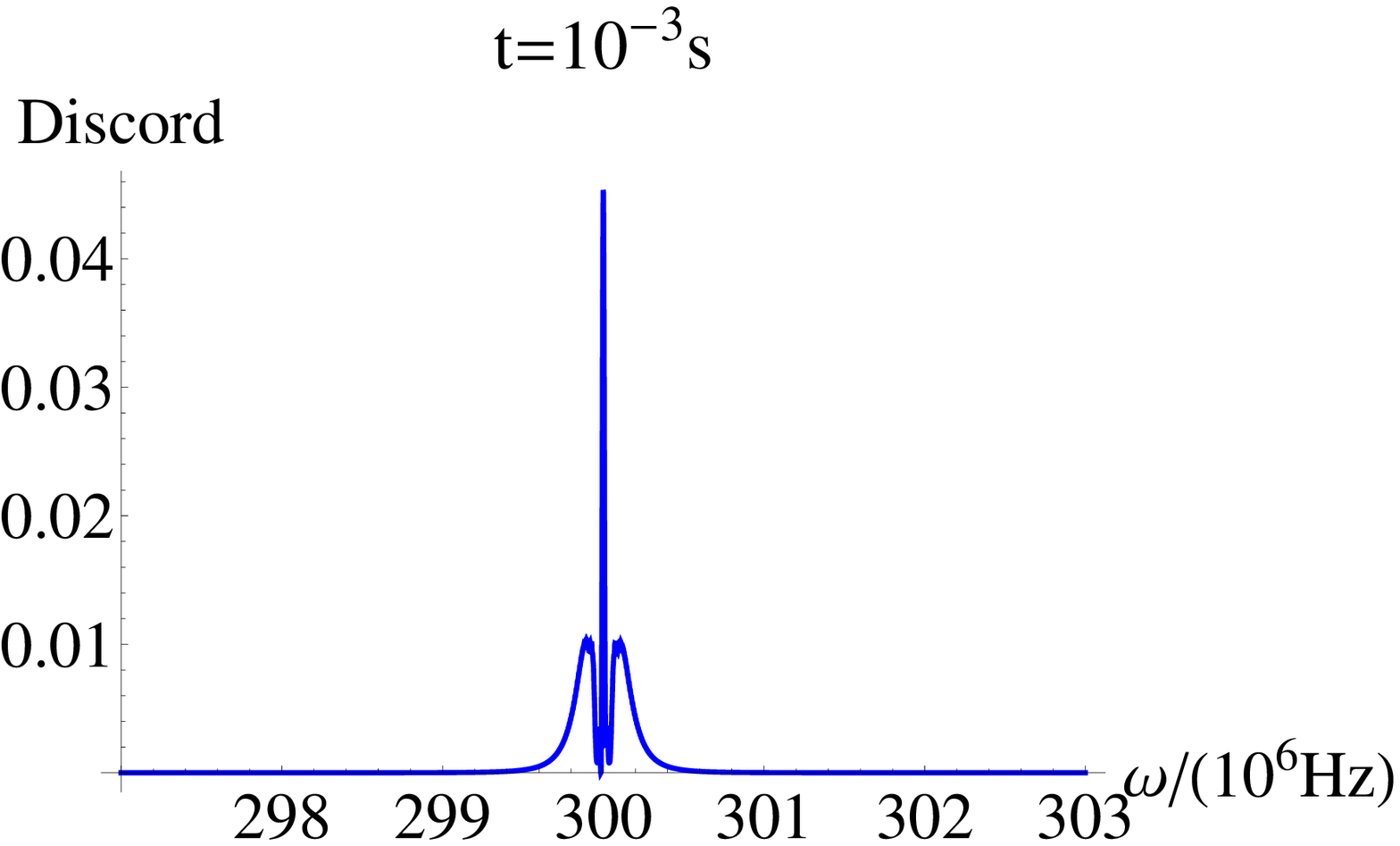} \
 \includegraphics[width=5cm]{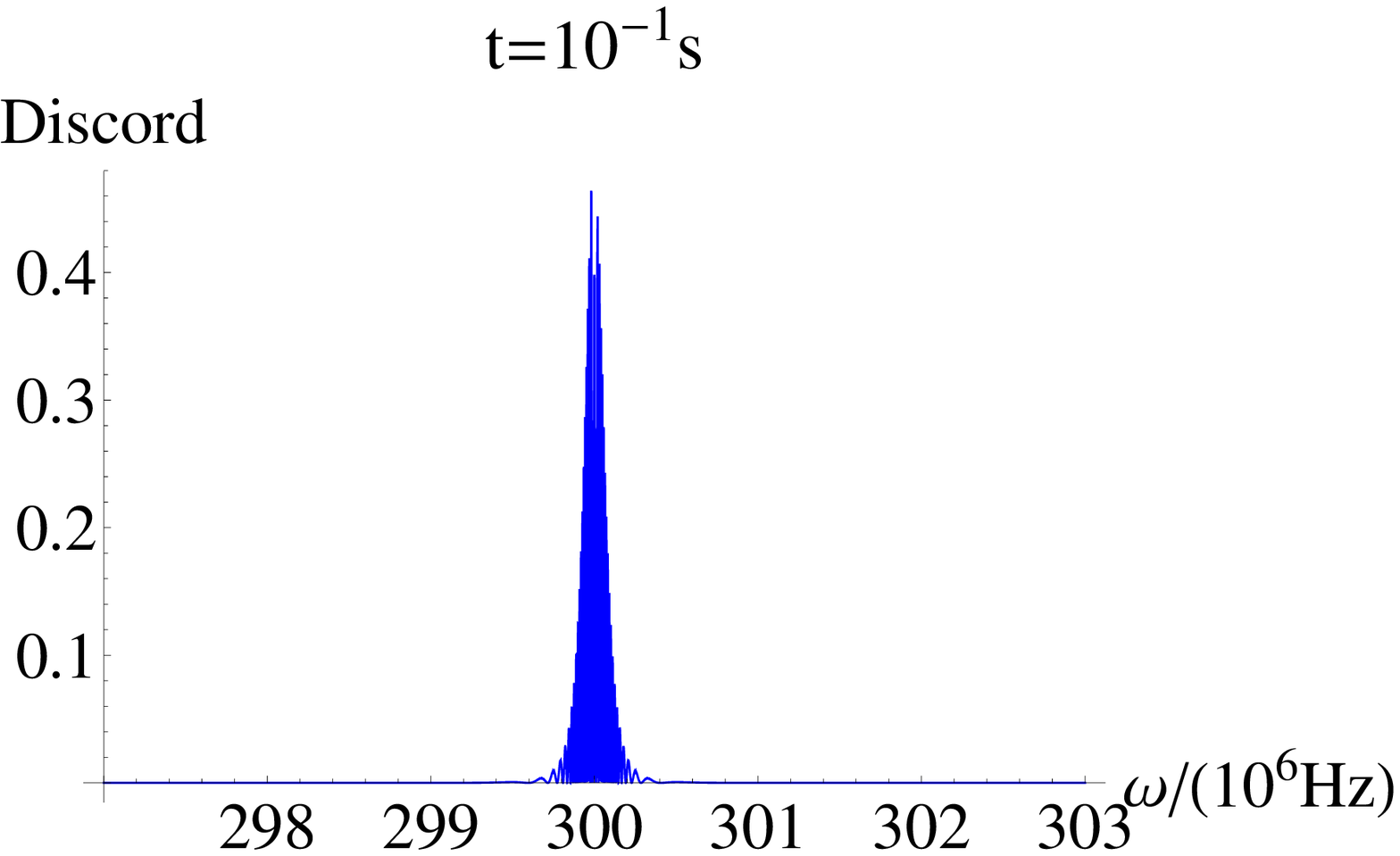}
 \caption{Pseudo-concurrence and quantum discord at
 time $t=10^{-5}s$, $t=10^{-3}s$, $t=10^{-1}s$.}
 \end{figure}

  \begin{figure}[!h]
  \includegraphics[width=8cm,height=5cm]{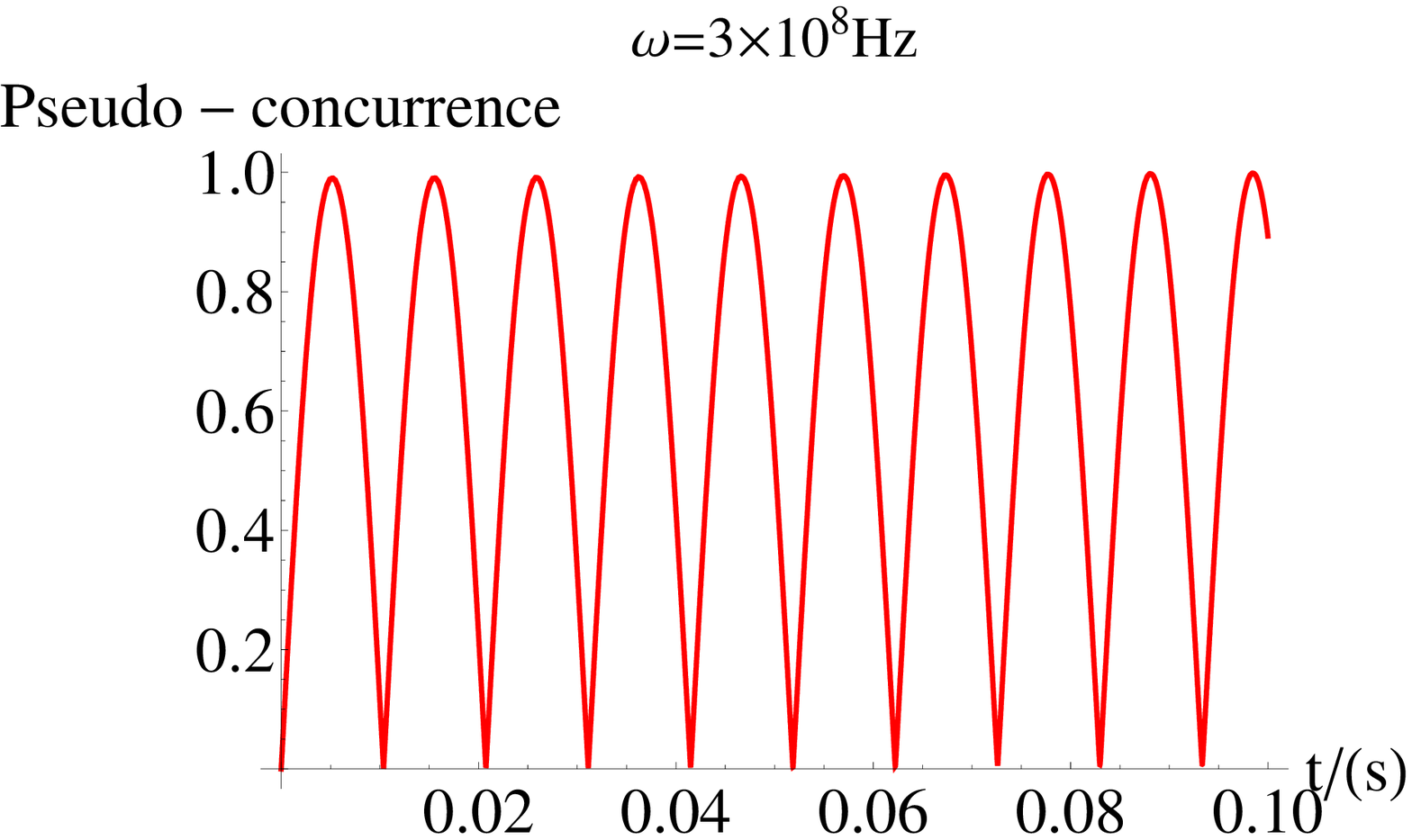}
  \includegraphics[width=8cm,height=5cm]{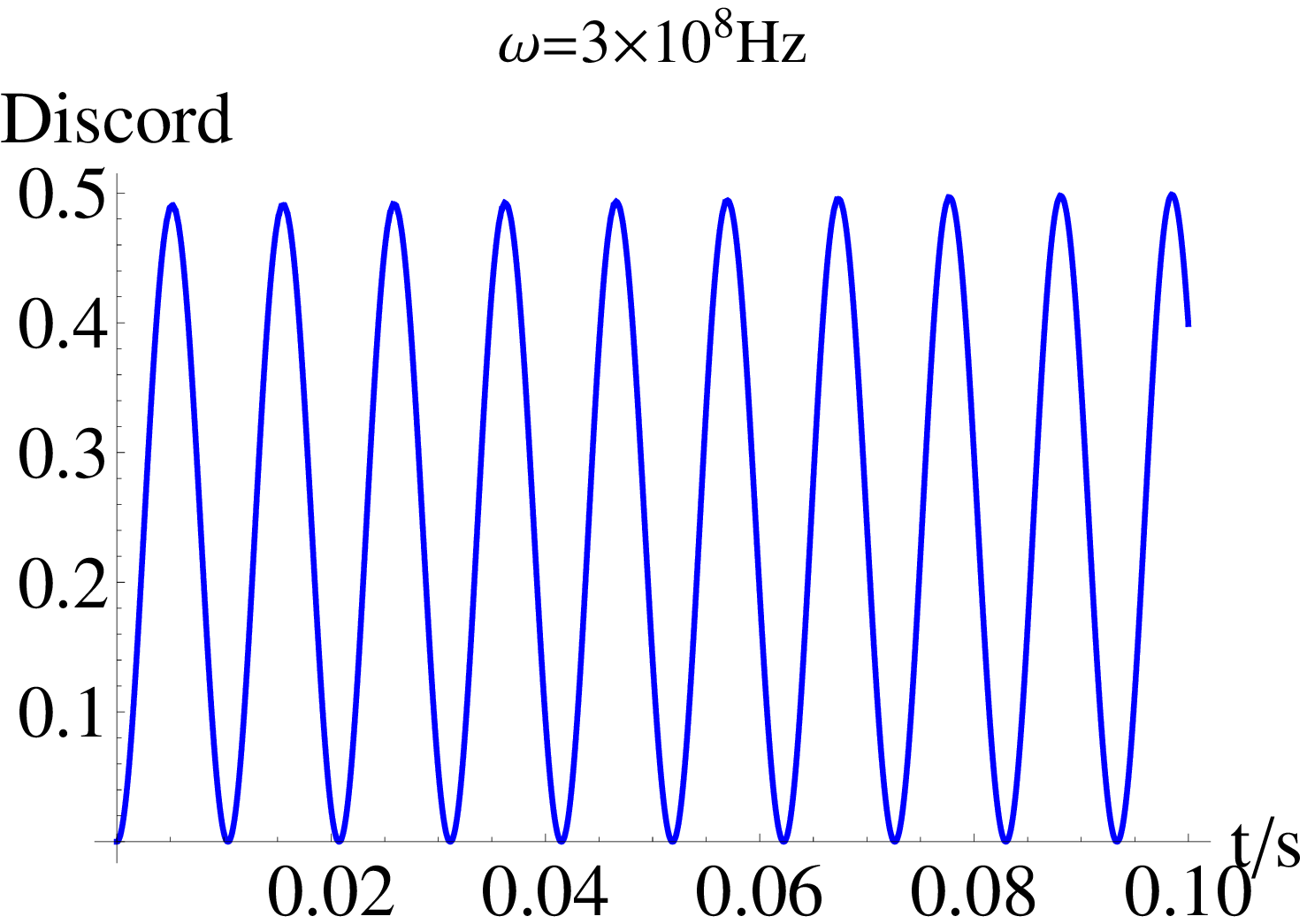}
  \caption{Pseudo-concurrence  and quantum discord at
 frequency $\omega=\frac{\omega _1+\omega _2}2=3\times 10^8Hz$.}
 \end{figure}

\section{Dynamics of entanglement and quantum discord of two qubits in NMR with
relaxation.}
When the evolution time gets  comparable to  the relaxation
time scales, the relaxation effects must be taken into account.
The relaxation process in NMR can be described in a phenomenological way
(\cite{Wijewardane2004};\cite{Oliveira2007}, 2.10)
\begin{eqnarray}
&&\frac d{dt}\rho (t)=\frac 1{i\hbar }[H(t),\rho (t)]-R, \\
&& \ \ \ \  R_{ij}=\frac{\rho _{ij}(t)-\rho _{ij}}{T_{ij}},  \\
&& \ \ \ \   T_{ij}=\delta _{ij}T_1+(1-\delta _{ij})T_2.
\end{eqnarray}
Where $T_1$ is the energy relaxation rate, $T_2$ is the phase randomization
rate. Theoretical calculations and experimental measurements for $T_1$ and $%
T_2$ are well developed. The typical value of $T_1$ is tens of seconds, and $T_2$ is easily on the order of 1 second or more.

Similar to Eq.(16), we let
\begin{eqnarray}
\rho (t)=\exp [\frac{i\omega t}2(\sigma _z\otimes I+I\otimes \sigma
_z)]\sigma (t)\exp [-\frac{i\omega t}2(\sigma _z\otimes I+I\otimes \sigma
_z)],
\end{eqnarray}
notice that
\begin{eqnarray}
\rho (0)=\sigma (0).
\end{eqnarray}
Using Eqs.(20-22), after some straightforward calculations, then Eq.(31) becomes
\begin{eqnarray}
\frac{d \sigma (t) }{dt}=\frac 1{i\hbar }[\widetilde{H},\sigma (t)]
 -\left(
\begin{array}{llll}
\frac{\sigma _{11}(t)}{T_1} & \frac{\sigma _{12}(t)}{T_2} & \frac{\sigma
_{13}(t)}{T_2} & \frac{\sigma _{14}(t)}{T_2} \\
\frac{\sigma _{21}(t)}{T_2} & \frac{\sigma _{22}(t)}{T_1} & \frac{\sigma
_{23}(t)}{T_2} & \frac{\sigma _{24}(t)}{T_2} \\
\frac{\sigma _{31}(t)}{T_2} & \frac{\sigma _{32}(t)}{T_2} & \frac{\sigma
_{33}(t)}{T_1} & \frac{\sigma _{34}(t)}{T_2} \\
\frac{\sigma _{41}(t)}{T_2} & \frac{\sigma _{42}(t)}{T_2} & \frac{\sigma
_{43}(t)}{T_2} & \frac{\sigma _{44}(t)}{T_1}
\end{array}
\right)+\left(
\begin{array}{llll}
\frac{\sigma _{11}(0)}{T_1} & \frac{\sigma _{12}(0)}{T_2}e^{-i\omega t} &
\frac{\sigma _{13}(0)}{T_2}e^{-i\omega t} & \frac{\sigma _{14}(0)}{T_2}%
e^{-2i\omega t} \\
\frac{\sigma _{21}(0)}{T_2}e^{i\omega t} & \frac{\sigma _{22}(0)}{T_1} &
\frac{\sigma _{23}(0)}{T_2} & \frac{\sigma _{24}(0)}{T_2}e^{-i\omega t} \\
\frac{\sigma _{31}(0)}{T_2}e^{i\omega t} & \frac{\sigma _{32}(0)}{T_2} &
\frac{\sigma _{33}(0)}{T_1} & \frac{\sigma _{34}(0)}{T_2}e^{-i\omega t} \\
\frac{\sigma _{41}(0)}{T_2}e^{2i\omega t} & \frac{\sigma _{42}(0)}{T_2}%
e^{i\omega t} & \frac{\sigma _{43}(0)}{T_2}e^{i\omega t} & \frac{\sigma
_{44}(0)}{T_1}
\end{array}
\right)
\end{eqnarray}
Eq.(36) is a first-order linear nonhomogeneous ordinary differential equations
with constant coefficients in the functions $\sigma _{11}(t),\sigma
_{12}(t),\sigma _{13}(t),\sigma _{14}(t),\sigma _{21}(t),...,\sigma _{44}(t)$%
. If we denote $\sigma (t)$ and the last matrix in Eq.(36) by
\begin{eqnarray}
&&\overrightarrow{\sigma }(t)=(\sigma _{11}(t),\sigma _{12}(t),\sigma
_{13}(t),\sigma _{14}(t),\sigma _{21}(t),...,\sigma _{44}(t))^t,  \\
&&\overrightarrow{f}(t)=(\frac{\sigma _{11}(0)}{T_1},\frac{\sigma _{12}(0)}{%
T_2}e^{-i\omega t},\frac{\sigma _{13}(0)}{T_2}e^{-i\omega t},\frac{\sigma
_{14}(0)}{T_2}e^{-2i\omega t},\frac{\sigma _{21}(0)}{T_2}e^{i\omega t},...,%
\frac{\sigma _{44}(0)}{T_1})^t,
\end{eqnarray}
then, Eq.(36) can be rewritten as a compact form
\begin{eqnarray}
\frac d{dt}\overrightarrow{\sigma }(t)=A\overrightarrow{\sigma }(t)+%
\overrightarrow{f}(t),
\end{eqnarray}
where, A is a $16\times 16$ matrix which is independent of t (the explicit expression
of A, see Appendix B). So the general solution of Eq.(39) is
\begin{eqnarray}
\overrightarrow{\sigma }(t)=\exp (tA)\overrightarrow{\sigma }%
(0)+\int_0^t\exp [(t-s)A]\overrightarrow{f}(t)ds.
\end{eqnarray}
We now take the zero initial state $|00\rangle $ for $\rho $, hence
\begin{eqnarray}
\overrightarrow{\sigma }(0)=(1,0,0,0,0,...,0)^t.  \\
\overrightarrow{f}(t)=(\frac 1{T_1},0,0,0,0,...,0)^t.
\end{eqnarray}
Then calculate Eqs.(40,34,2,10), we will get $C(\rho (t))$ and $D_A^G(\rho
(t))$.
From the discussions about Fig.1 and Fig.2,  in Fig.3 below we only discuss the dynamics of $\overline{C}(\rho (t))$ and $D_A^G(\rho
(t))$ at frequency $\frac{\omega _1+\omega _2}2=3\times 10^8Hz$, and we take the parameters of NMR as in Fig.1.

 \begin{figure}[!h]
 \includegraphics[width=16cm]{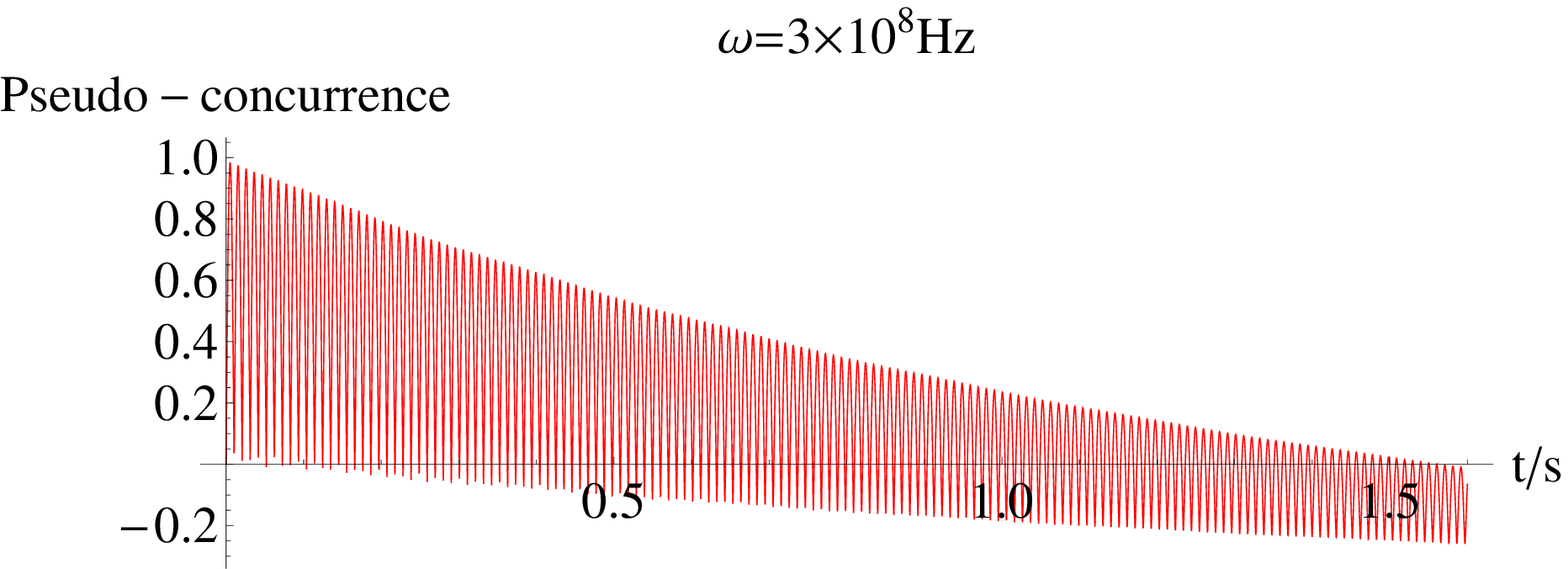}
 \includegraphics[width=16cm,trim=-45 0 0 0]{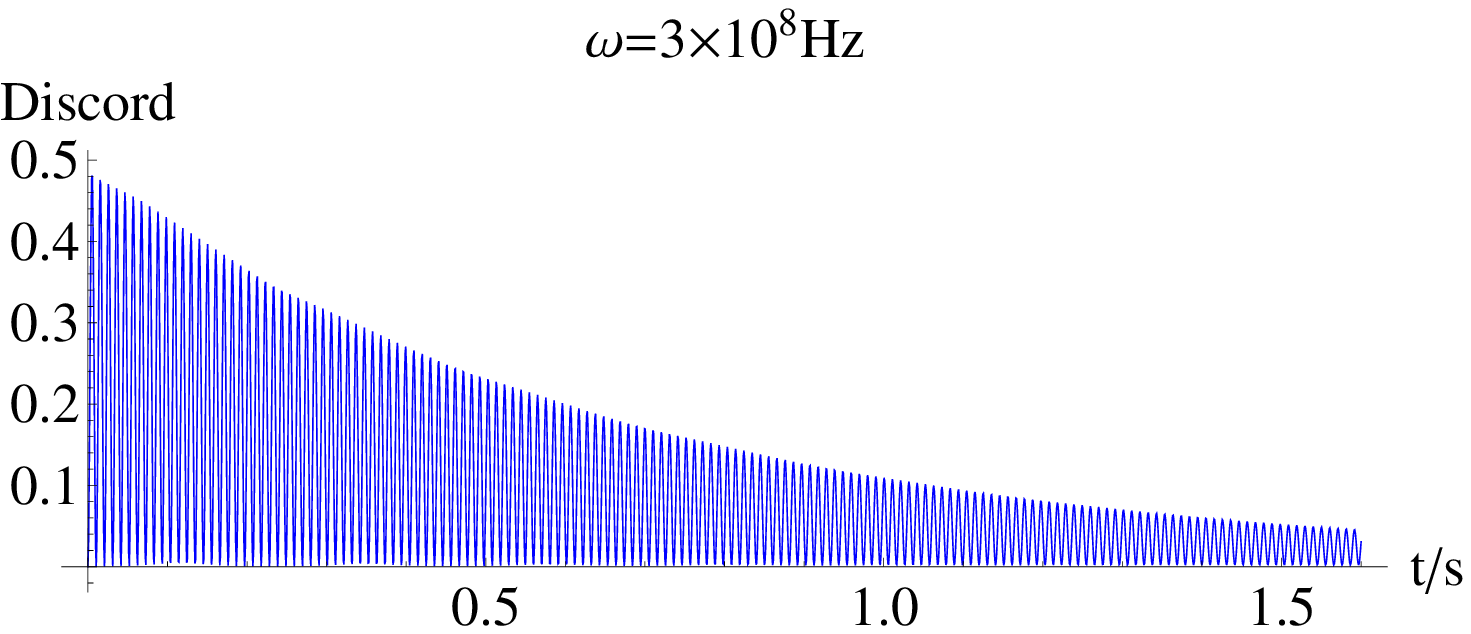}
 \caption{Pseudo-concurrence and quantum discord  at
 frequency $\omega=\frac{\omega _1+\omega _2}2=3\times 10^8Hz$.}
 \end{figure}

From Fig.3 we see that, when the evolution time is much shorter than the
relaxation scales, the dynamics of entanglement and quantum discord are
similar. But for a longer time, namely, near relaxation scales, the
entanglement manifests a sequence of sudden deaths and revivals, and finally
disappears completely. At the same time, the quantum discord, after a
sequence of oscillations, still retains remarkable values.

In a dynamical process, the phenomenon of entanglement disappears in a finite time  is named entanglement sudden death \cite{Yu2004}. This is an interesting phenomenon in contrast to the usual intuition that decoherence will be  in  an infinite time. Also, this is an important phenomenon for quantum information processing since much of quantum information processing rely on entanglement \cite{Nielson2000,Horodecki2009}.

\section{Summary}

In summary, we investigated the dynamics of entanglement and quantum discord of two qubits in liquid state homonuclear NMR systems.
We showed that the dynamical behaviors of entanglement and quantum discord are
similar in short time where relaxation effects can be neglected. When the time is
long enough to be comparable to the relaxation rates, the entanglement manifests
the phenomenon of sudden deaths and survivals, and at last disappears completely. Meanwhile the quantum discord retains remarkable
values. That is to say, quantum discord is more robust than
entanglement against the relaxation processes. Hence quantum algorithms based only on quantum discord correlation may be more robust than
those based on entanglement.

\section*{Acknowledgements}

This work was supported by National Natural Science Foundation of China
(Grant Nos. 10775101). The authors thank Qing Hou and Bo You for helpful discussions.

 \appendix
 \section{Explicit expressions for $\{x_i\}_{i=1}^3$
and $\{T_{ij}\}_{i,j=1}^3$ by the elements of $\rho $ in Eq.(11)}
$x_1=\rho _{13}+\rho _{24}+\rho _{31}+\rho _{42}$

$x_2=i(\rho _{13}+\rho _{24}-\rho _{31}-\rho _{42})$

$x_3=\rho _{11}+\rho _{22}-\rho _{33}-\rho _{44}$

$T_{11}=\rho _{14}+\rho _{23}+\rho _{32}+\rho _{41}$

$T_{12}=i(\rho _{14}-\rho _{23}+\rho _{32}-\rho _{41})$

$T_{13}=\rho _{13}-\rho _{24}+\rho _{31}-\rho _{42}$

$T_{21}=i(\rho _{14}+\rho _{23}-\rho _{32}-\rho _{41})$

$T_{22}=-\rho _{14}+\rho _{23}+\rho _{32}-\rho _{41}$

$T_{23}=i(\rho _{13}-\rho _{24}-\rho _{31}+\rho _{42})$

$T_{31}=\rho _{12}+\rho _{21}-\rho _{34}-\rho _{43}$

$T_{32}=i(\rho _{12}-\rho _{21}-\rho _{34}+\rho _{43})$

$T_{33}=\rho _{11}-\rho _{22}-\rho _{33}+\rho _{44}$

\section{Explicit expression of matrix A in Eq.(39)}

For simplicity, we put
$\frac{\omega _1+\omega _2}2=a$,
$\frac{\omega _1-\omega _2}2=b$,
$J=c$,
$\frac{g_1}2=d$,
$\frac{g_2}2=e$,
then

$A=i\{\{0,-e,-d,0,e,0,0,0,d,0,0,0,0,0,0,0\},$

$\ \ \ \ \ \ \ \ \ \{-e,a-y-b-2c,0,-d,0,e,0,0,0,d,0,0,0,0,0,0\},$

$\ \ \ \ \ \ \ \ \ \{-d,0,a-y+b-2c,-e,0,0,e,0,0,0,d,0,0,0,0,0\},$

$\ \ \ \ \ \ \ \ \ \{0,-d,-e,2a-2y,0,0,0,e,0,0,0,d,0,0,0,0\},$

$\ \ \ \ \ \ \ \ \ \{e,0,0,0,-a+y+b+2c,-e,-d,0,0,0,0,0,d,0,0,0\},$

$\ \ \ \ \ \ \ \ \ \{0,e,0,0,-e,0,0,-d,0,0,0,0,0,d,0,0\},$

$\ \ \ \ \ \ \ \ \ \{0,0,e,0,-d,0,2b,-e,0,0,0,0,0,0,d,0\},$

$\ \ \ \ \ \ \ \ \ \{0,0,0,e,0,-d,-e,a-y+b+2c,0,0,0,0,0,0,0,d\},$

$\ \ \ \ \ \ \ \ \ \{d,0,0,0,0,0,0,0,-a+y-b+2c,-e,-d,0,e,0,0,0\},$

$\ \ \ \ \ \ \ \ \ \{0,d,0,0,0,0,0,0,-e,-2b,0,-d,0,e,0,0\},$

$\ \ \ \ \ \ \ \ \ \{0,0,d,0,0,0,0,0,-d,0,0,-e,0,0,e,0\},$

$\ \ \ \ \ \ \ \ \ \{0,0,0,d,0,0,0,0,0,-d,-e,a-b+2c-y,0,0,0,e\},$

$\ \ \ \ \ \ \ \ \ \{0,0,0,0,d,0,0,0,e,0,0,0,-2a+2y,-e,-d,0\},$

$\ \ \ \ \ \ \ \ \ \{0,0,0,0,0,d,0,0,0,e,0,0,-e,-a-b-2c+y,0,-d\},$

$\ \ \ \ \ \ \ \ \ \{0,0,0,0,0,0,d,0,0,0,e,0,-d,0,-a+b-2c+y,-e\},$

$\ \ \ \ \ \ \ \ \ \{0,0,0,0,0,0,0,d,0,0,0,e,0,-d,-e,0\}\}$

$\ \ \ \ \ -DiagonalMatrix[\{\frac 1{T_1},\frac 1{T_2},\frac 1{T_2},\frac 1{T_2},\frac 1{T_2},\frac 1{T_1},\frac 1{T_2},\frac 1{T_2},\frac 1{T_2},\frac 1{T_2},\frac 1{T_1},\frac 1{T_2},\frac 1{T_2},\frac 1{T_2},\frac 1{T_2},\frac 1{T_1}\}].$

\bibliographystyle{model1-num-names}

\end{document}